# On the thermal rectification factor in steady heat conduction

Hamou Sadat and Vital Le Dez

*Institut P', Université de Poitiers, Centre National de la Recherche Scientifique, 2 Rue Pierre Brousse, Bâtiment B25, 86073 Poitiers Cedex 9, France*

*Corresponding author hamou.sadat@univ-poitiers.fr
Tel.:+33-05-49-45-38-10; fax: +33-05-49-45-35-45

## Abstract

Thermal rectification in heat conduction problems has been extensively studied in planar slabs. Here we consider the rectification problem in planar, cylindrical and spherical geometries involving two layers one of which has a temperature variable heat conductivity. The rectification factor is analytically calculated. It is shown that a maximum theoretical value of 1.618 is obtained.

*Keywords:* Rectification factor, thermal diode, heat conduction

## 1. Introduction

In many heat transfer processes, it is necessary to transfer the flow of heat in a desired direction but block the heat flow in the opposite direction. Heat transfer devices with this feature are termed thermal diodes analogous to the electrical diodes which are commonly employed for rectification in electrical circuits. This phenomenon can appear in heat conduction [1-2], heat convection [3-4] and heat radiation [5-7] problems as well. The idea of heat conduction thermal rectifier in a composite wall consisting of two materials, each having temperature dependent conductivity was first presented in [8-9]. More recently an interesting study [10] has shown that the maximum rectification factor in a slab is equal to 3 when the two layers have variable heat conductivities. In this note we consider the case of two materials one of which has a linear increase of its conductivity with temperature while the other has a constant conductivity and we consider the planar, spherical and cylindrical geometries. The steady heat conduction problem is solved analytically and an expression for the rectification factor is given. It is shown that a maximum theoretical value of 1.61 is obtained.

## 2. Forward case in the spherical shell

Let us consider a bilayer spherical shell as indicated on Fig.1. The inner layer has a temperature dependent thermal conductivity while the conductivity of the outer layer is supposed to be constant. Furthermore, the temperature dependence of the conductivity is supposed to be linear:

$$\lambda_1(T) = \lambda_0(1 + \alpha T) \quad (1)$$

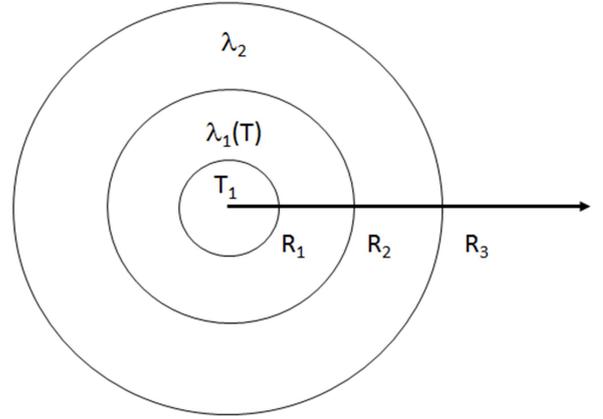

Figure 1: Bilayer spherical shell

We suppose also that the boundary conditions are constant temperature $T_1$ at $r=R_1$ and a zero temperature at $r=R_3$ :

$$T = T_1 \quad at \quad r = R_1 \quad (2\text{-a})$$
$$T = 0 \quad at \quad r = R_3 \quad (2\text{-b})$$

The case of non-zero temperature at $r=R_3$ can be treated by a simple variable change. At steady state the heat equation writes:

$$\frac{1}{r^2}\frac{d}{dr}\left(r^2 \lambda(T)\frac{dT}{dr}\right) = 0 \quad (3)$$

By solving this equation in the two layers and using the flux continuity condition at $r=R_2$, one can easily show (as done in the annexe) that the interface temperature obeys the following second degree equation:

$$\frac{\alpha}{2}T_f^2 + T_f[1 + ks] - T_1 - \frac{\alpha}{2}T_1^2 = 0 \quad (4)$$

Where $T_f$ stands for the interface temperature in this forward case and where $k = \frac{\lambda_2}{\lambda_0}$ and $s = \frac{R_3}{R_1}\frac{(R_2-R_1)}{(R_3-R_2)}$. The global heat conduction flux then writes:

$$Q_f = \frac{4\pi\lambda_2 R_3 R_2 T_f}{(R_3-R_2)} \quad (5)$$

## 3. Reverse case

Let us now consider the reverse case where temperature $T_1$ is imposed at the outer boundary while the inner boundary is set to T=0. One then can show that the solution to the heat equation leads to the following interface temperature where Tr stands to the reverse interface temperature and where λ and s are defined as previously:

$$\frac{\alpha}{2}T_r^2 + T_r[1+ks] - ksT_1 = 0 \quad (6)$$

The global flux in this reverse case is therefore given by:

$$Q_r = 4\pi\lambda_2 R_3 R_2 \frac{(T_1-T_r)}{(R_3-R_2)} \quad (7)$$

## 4. Rectification factor

The rectification factor is usually defined by the ratio of the heat fluxes: $R = \frac{Q_f}{Q_r}$. From equations (5) and (7), it comes:

$$R = \frac{T_f}{T_1-T_r} \quad (8)$$

By introducing the flux expressions and noting that: $(1+\alpha T_1) = \frac{\lambda_1}{\lambda_0}$, one finally obtains the following expression :

$$R = \frac{-(1+ks)+\sqrt{(1+ks)^2+\left[\left(\frac{\lambda_1}{\lambda_0}\right)^2-1\right]}}{\left(\frac{\lambda_1}{\lambda_0}+ks\right)-\sqrt{(1+ks)^2+2ks\left(\frac{\lambda_1}{\lambda_0}-1\right)}} \quad (9)$$

## 5. Cylindrical shell and planar slab

By using the same approach and the same boundary conditions as in the previous case, the rectification factor for the cylindrical shell and the planar slab are found to be expressed by relation (9) but with $s = \frac{\log\left(\frac{R_2}{R_1}\right)}{\log\left(\frac{R_3}{R_2}\right)}$ and s=1 for the cylindrical and the slab cases respectively. This rectification factor R is sketched on Figure 2 for different values of the two ratios $j = \frac{\lambda_1}{\lambda_0}$ and $k = \frac{\lambda_2}{\lambda_0}$ in the case of the cylindrical geometry with the radius values: $R_1$=0.15, $R_2$=0.17 and $R_3$=0.19. One can see that there exist a region where the R value is near 1.6 or even greater. Similar figures and conclusions can be drawn for the slab and spherical cases.

## 6. Maximum value of the rectification factor

Our numerous simulations showed that the great values of R are obtained for high values of j. This is due to the fact that the rectification is a result of the non linearity. One can then expect that the maximum rectification factor would be obtained for the highest values of the ratio: $j = \frac{\lambda_1}{\lambda_0}$. If we now express R as a function of $x=k/j=\frac{\lambda_2}{\lambda_1}$ and by taking the limit when j tends to infinity, one can see that:

$$R \to \frac{-x+\sqrt{1+x^2}}{(1+x)-\sqrt{x^2+2x}} \quad (10)$$

This function is plotted on Figure 3. There exist a maximum which can be calculated by writing: $\frac{dR}{dx} = 0$. It is easily shown that this is obtained for $x = \frac{1}{2}$ and that the maximum value of the rectification factor is equal to:

$$R_M = \frac{1+\sqrt{5}}{2} = 1.618 \quad (11)$$

This maximum value is surprisingly equal to the golden ratio and is to be compared to the maximum value of 3 obtained in [10].

## 7. Conclusion

In this study we considered steady heat conduction in a bilayer walls of planar, cylindrical and spherical geometry. A diode effect has been identified when the heat conductivity of one layer varies with temperature while the other one remains constant. Simple analytical expression for the rectification factor has been given. Theoretical maximum and minimum values have also been discussed.

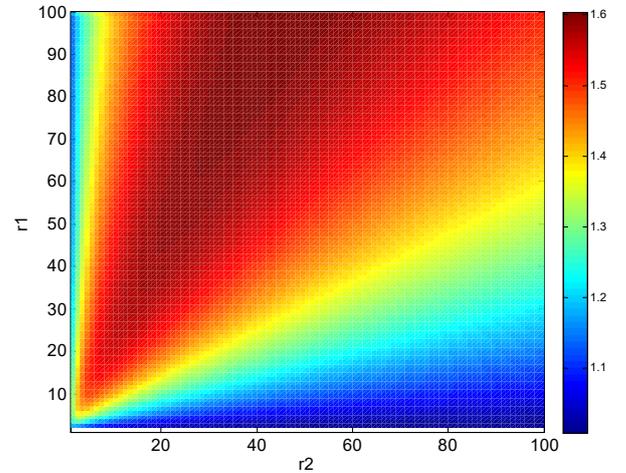

Figure 2: Rectification factor for the cylindrical geometry: $R_1$=0.15; $R_2$=0.17; $R_3$=0.19;

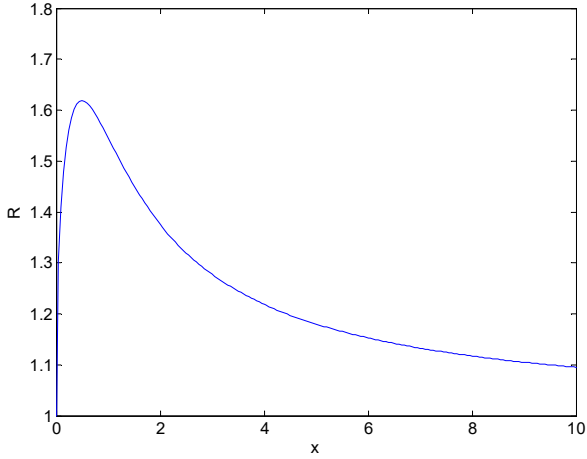

Figure 3: Rectification factor as a function of x

# References

[1] Baowen Li, Lei Wang, and Giulio Casati, Thermal Diode: Rectification of Heat Flux, Physical Review Letters, Vol. 93, N°18, 2004.

[2] B. Hu, D.HE, L. Yang and Y. Zhang, Asymmetric heat conduction through a weak link, Physical review E,74,060201, 2006.

[3] N.Seki, S.Fukusako, A.Yamaguchi, An experimental study of free convective heat transfer in a parallelogrammic enclosure, Journal of Heat Transfer, Vol.105, August 1983.

[4] H. Nakamura and Y. Asako, Heat transfer in a parallelogram shaped enclosure, Bulletin of the JSME, vol.23, N°185, November, 1980.

[5] S.A. Biehs, D. Reddig and M. Holthaus, Thermal radiation and near-field energy density of thin metallic films, European Physical Journal B, 55, 3, 2007, 237-251.

[6] Philippe Ben-Abdallah and Svend-Age Biehs, Phase-change radiative thermal diode, Applied Physics Letters, 103, 191907, 2013.

[7] M. Terraneo, M. Peyrard and G. Casati, Controlling the energy flow in nonlinear lattices: a model for a thermal rectifier, Phys. Rev. Lett., Vol. 88, N°9 ,094302-1-4, 2002.

[8] M.Peyrard, The design of a thermal rectifier, Europhysics letters, 76, 49, 2006.

[10] Tien-Mo Shih, Zhaojing Gao, Ziquan Guo, Holger Merlitz, Patrick J. Pagni and Zhong Chen, Maximal rectification ratios for idealized bi-segment thermal rectifiers, Scientific Reports, 5:12677, DOi: 10.1038/srep12677

# Annexe

In the second region where $\lambda_2$ is constant, one has:

$$\lambda_2 \frac{dT}{dr} = \frac{C}{r^2} \quad (A1)$$

Solving this equation with the condition $T(R_3)=0$ leads to:

$$T = \frac{C}{\lambda_2}\left(\frac{1}{R_3} - \frac{1}{r}\right) \quad (A2)$$

If $T_f$ is the interface temperature, constant C can now be written as:

$$C = \frac{\lambda_2 R_2 R_3 T_f}{(R_2 - R_3)} \quad (A3)$$

The flux at the interface is therefore:

$$q(R_2) = \frac{\lambda_2 R_3 T_f}{R_2(R_3 - R_2)} \quad (A4)$$

In the first region, the following equation holds:

$$\lambda_0 (1 + \alpha T) \frac{dT}{dr} = \frac{A}{r^2} \quad (A5)$$

Integrating the equation between R1 and R2, one obtains:

$$\int_{T_1}^{T_f}(1 + \alpha T)dT = \int_{R_1}^{R_2} \frac{A\,dr}{\lambda_0 r^2} \quad (A6)$$

Hence:

$$(T_f - T_1) + \frac{\alpha}{2}(T_f^2 - T_1^2) = \frac{A}{\lambda_0} \frac{(R_2 - R_1)}{R_2 R_1} \quad (A7)$$

The flux at the interface is:

$$q = -\frac{A}{R_2^2} \quad (A8)$$

Equating A8 and A4, one gets finally:

$$A = \frac{\lambda_2 R_3 R_2 T_f}{R_3 - R_2} \quad (A9)$$

Inserting this form in A7 leads to the desired equations (4-5). The reverse case can be treated in a similar way.